\documentclass[
 reprint,
 amsmath,amssymb,
 aps,
 pra,
]{revtex4-2}

\usepackage{graphicx}% Include figure files
\usepackage{dcolumn}% Align table columns on decimal point
\usepackage{bm}% bold math
\usepackage{amsmath}
\usepackage{tikz}
\usepackage{dsfont}
\begin{document}

\preprint{APS/123-QED}

\title{Quantum search in sets with prior knowledge}% Force line breaks with \\

\author{Umut Çalıkyılmaz}
 \email{umutc@metu.edu.tr}
\author{Sadi Turgut}%
 \email{sturgut@metu.edu.tr}
\affiliation{%
 Middle East Technical University, Physics Department, Dumlupınar Bulvarı 06800 Çankaya Ankara, Turkey\\
}%

\date{\today}% It is always \today, today,

\begin{abstract}
Quantum Search Algorithm made a big impact by being able to solve the search problem for a set with $N$ elements using only $O(\sqrt{N})$ steps. Unfortunately, it is impossible to reduce the order of the complexity of this problem, however, it is possible to make improvements by a constant factor. In this paper we pursued such improvements for search problem in sets with known probability distributions. We have shown that by using a modified version of quantum search algorithm, it is possible to decrease the expected number of iterations for such sets.
\end{abstract}

\pacs{Valid PACS appear here}% PACS, the Physics and Astronomy
\maketitle

%\tableofcontents

\section{\label{sec:level1}Introduction}

As in any other advancement in science, the idea of quantum computing has emerged from need. Classical computing is insufficient for simulating complex quantum systems mainly because the memory needed to store the state of a quantum system as classical information goes exponentially with the system size. As a better way of simulating this kind of systems, Richard Feynman proposed the use of quantum computers, computers employing quantum systems to store and process data \cite{ref1, ref2}. Shortly after, other advantages of using such information processing method are noticed. First, the superiority of quantum computing over classical computing is shown using some specially designed problems \cite{ref3, ref4, ref5, ref6}. Then Shor proved the possibility of solving the age-old factorization problem in polynomial time by using a quantum computer \cite{ref7}.

A few years after that, Grover showed that another classical problem, namely the search problem, can be solved in a shorter time using a quantum algorithm \cite{ref8, ref9}. The search problem is a problem of finding an element, satisfying some condition, inside an unordered set. It is solved classically by trying every element in the set until a solution is found. Grover's algorithm works by applying successive rotations on some initial quantum state until it is transformed into a desired state. The initial state is the superposition of all elements in the set with equal coefficients, and the desired final state is the superposition of only the solutions.

Grover's algorithm is not able to solve the search problem in a polynomial time, but it reduces the number of required trials dramatically. In case of a single solution, which is the only case that we have examined in this paper, searching a set of $N$ elements classically takes $O(N)$ trials. Grover's algorithm finds the solution for the same problem in only $O(\sqrt{N})$ iterations of the Grover operator.

It has been shown that the order of the number of iterations required to solve this problem using a quantum algorithm cannot be smaller than $O(\sqrt{N})$ \cite{ref10}. However, some improvement to the Grover's algorithm by a constant factor is still possible. The exact Grover algorithm requires the repetition of Grover operator $\pi/4\sqrt{N}$ times. In Ref.~\onlinecite{ref11}, a different method, where the Grover operator is repeated by a smaller number of times, is introduced. As finding a solution is not guaranteed at the measurement stage, it might be necessary to repeat the attempt several times until the solution is found. With this approach, the average time to find the solution can be reduced by $12\,\%$.

The aim of this paper is achieving a similar reduction in the case that the elements of the search space has differing probabilities of being a solution. It is assumed that for each element $i$, there is a known probability $p_i$, which denotes the probability of element $i$ being the solution. To achieve a better result, different initial states than that of Grover's method are utilized. In these states, coefficients of different indices can be different from each other. Here, the cost of preparing these states are neglected. The method using these generic initial states is called as the generic search.

In practical applications of search algorithms, it is sometimes possible to assume differing probabilities for the elements of the search set. For example, when brute force is used to crack a password, some combinations are assumed to be more probable and being tried before others. Quantum search can be utilized for such a task, and generic search, or any similar method using generic initial states, can be used to speed up the process. 

In the next section of this paper, the procedure and the parameters of generic search is defined and various equations are derived. The third section is reserved for finding the optimality conditions for the parameters of the method. In the fourth section, the estimated values for the optimum values of the parameters and the expected number of steps are presented. The fifth section concludes the paper by summarizing our study and giving some ideas about the implementation of the generic search.

\section{\label{sec:level2}Definition of the Method}

\subsection{The Procedure and the Parameters of the Method \label{subsec:2.1}}

Generic search uses consecutive quantum searches until a solution to the search problem is found, such as the method given in \cite{ref11}. The difference of generic search is that it employs varying initial quantum states for the searches. Thus, we expect that the optimal number of applied Grover iterations would also be different for each step.

The $j^{th}$ quantum search in this method is named as the $j^{th}$ step of the algorithm. The initial state used for the $j^{th}$ step is denoted as $\left|\psi^{(j)} \right\rangle$ and defined as

\begin{equation}
	\left|\psi^{(j)} \right\rangle=\sum_{i=0}^{N}c_i^{(j)} \left|i \right\rangle ~,
	\label{initstate}
\end{equation}
where $c_i$ are assumed to be real and positive for convenience. At the $j^{th}$ step, $m^{(j)}$ iterations are applied to this initial state before measuring the final state. As one might expect, $m^{(j)}<\pi/4\sqrt{N}$ for each $j$, where $N$ is the number of elements in the search space.

We followed a convention similar to the one used in \cite{ref12} to define the parameters of the algorithm. Let $s$ denote the index of the solution that we sought after. Then, the state representing the solution is $\left|s \right\rangle$. The non-solution state for step-$j$ $\left|ns^{(j)} \right\rangle$ is defined as
\begin{equation}
    \left\vert ns^{(j)} \right\rangle=\frac{1}{\sqrt{1-\left\vert c^{(j)}_s\right\vert^2}} \sum_{i \neq s} c^{(j)}_i\vert i\rangle ~.
\end{equation}

Similar to the original Grover algorithm, the oracle operator $O$ reflects a state along $\vert s\rangle$,
\begin{equation}
    O = \mathds{1}-2\vert s\rangle\langle s\vert ~,
\end{equation}
and the succeeding reflection operator for the $j$th step is
\begin{equation}
    R^{(j)} = \mathds{1}-2\left\vert\psi^{(j)} \right\rangle \left\langle \psi^{(j)} \right\vert ~.
\end{equation}
Note that if the state $\vert\psi^{(j)}\rangle$ can be prepared easily on a quantum computer, then the operation $R^{(j)}$ can also be implemented easily\footnote{If $G_j$ is the unitary map corresponding to the gates that prepares the state $\vert\psi^{(j)}\rangle$ from the standard state, $\vert\psi^{(j)}\rangle=G_j\vert0\vert$, then $R^{(j)}=G_j(\mathds{1}-2\vert0\rangle\langle0\vert)G_j^\dagger$.}. We will assume this for all $\vert\psi^{(j)}\rangle$ and ignore the complexity of state preparation.

Let $\left| \xi \right\rangle$ be a quantum state lying in the plane spanned by $\left|s \right\rangle$ and $\left|ns^{(j)} \right\rangle$. As it was defined in the Grover algorithm, the Oracle operator reflects a state with respect to the subspace spanned by all the vectors of non-solution indices. When applied to $\left|\xi \right\rangle$, $O$ reflects this state with respect to $\left|ns^{(j)} \right\rangle$ and the reflected state $\left|\xi' \right\rangle$ would still be in the plane. Then, if $R^{(j)}$ is applied to $\left|\xi' \right\rangle$, the resulting state $\left|\xi'' \right\rangle$ is also in the plane. The reason is that $\left|\psi^{(j)} \right\rangle$ is a superposition of $\left|s \right\rangle$ and $\left|ns^{(j)} \right\rangle$, which can be seen from Eq.~\eqref{initstate}, so it lies in the plane spanned by these two states. Moreover, being a product of two reflections, the operator $G^{(j)}=R^{(j)}O$ acts as a rotation for $\left|\xi \right\rangle$. The angle of rotation is $2\phi^{(j)}_s$ where $\phi^{(j)}_s$ is the angle between $\left|\psi^{(j)} \right\rangle$ and $\left|ns^{(j)} \right\rangle$,
\begin{equation}
    \phi^{(j)}_s=\arcsin\left| \left\langle \psi^{(j)}|s \right\rangle  \right| = \arcsin c_s^{(j)} ~.
\end{equation}
In step $j$, the operator $G^{(j)}$ is applied $m^{(j)}$ times to the initial state $\left|\psi^{(j)} \right\rangle$. Then, the angle between the final state of step $j$ and $\left|ns^{(j)} \right\rangle$ is 
\begin{equation}
    \theta^{(j)}_s=(2m^{(j)}+1)\phi^{(j)}_s ~.
\end{equation}

One important point is that the index of the solution is not known prior to the search. Thus, defining the final angle depending on $s$ is not useful. Instead, the hypothetical final angle in the case that index $i$ is the solution must be found, for each $i$. This parameter can be defined as
\begin{equation}
	\theta^{(j)}_i= (2m^{(j)}+1) \arcsin c_i^{(j)} ~.
	\label{restoreci}
\end{equation}

Let $P^{(j)}(\mathrm{succ}|i)$ denote the probability that the correct solution would be identified successfully if $i$ were the solution (if $s=i$). Then 
\begin{equation}
    \begin{aligned}
        P^{(j)}(\mathrm{succ}|i)&=\left| \left\langle i \left| \left(G^{(j)}\right)^{m^{(j)}} \right| \psi_j \right\rangle  \right|^2 \\
         &=\sin^2\theta^{(j)}_i ~, 
        \end{aligned}
	\label{succprob}
\end{equation}
and 
\begin{equation}
    \begin{aligned}
        P^{(j)}(\mathrm{fail}|i)&=1-P^{(j)}(\mathrm{succ}|i) \\
        &=\cos^2\theta^{(j)}_i ~. 
    \end{aligned}
\end{equation}
In the rest of this paper, the final angles $\theta^{(j)}_i$ are used as primary variables for convenience. Once the $\theta^{(j)}_i$ values are known, the initial angles and the coefficients can be calculated easily using Eq.~\eqref{restoreci}.

\subsection{Expected Number of Iterations \label{subsec:2.2}}

The purpose of generic search is to reduce the expected number of Grover iterations required to solve the search problem. To find the minimum possible value for this variable, first, its mathematical expression must be constructed. As a first step, assume that the solution is the index $i$ and let $E_i$ be the expected number of iterations in this case. It is defined as
\begin{equation}
    \begin{aligned}
        E_i&=m^{(1)}+\cos^2\theta_i^{(1)}m^{(2)}+... \\
        &=\sum_{j\geq1}m^{(j)}\prod_{k=1}^{j-1}\cos^2\theta_i^{(k)} ~, 
    \end{aligned}
	\label{ei}
\end{equation}
where we follow the convention that when $j=1$, the product in the expression above is equal to $1$, i.e.,

\begin{equation}
	\prod_{k=1}^{0}\cos^2\theta_i^{(k)}=1 ~.
\end{equation}

In Eq.~\eqref{ei}, each term in the sum is the number of iterations applied in the $j^{th}$ step times the probability of ever carrying out the $j^{th}$ step. For instance, the first quantum search is applied definitely, so the first term of the sum is $m^{(1)}$. Likewise, the second search is applied in the case that the solution is not found in the first step, with a probability of $\cos^2 \theta_i^{(1)}$, so the second term in the sum is $\cos^2 \theta_i^{(1)} m^{(2)}$.

Since the solution is not known initially, writing the expected number of iterations in a more general manner is required. This can be done using $E_i$ values defined in Eq.~\eqref{ei} and the probabilities $p_i$ as
\begin{equation}
	E=\sum_{i=0}^{N-1} p_iE_i=\sum_{i=0}^{N-1} p_i \sum_{j\geq1}m^{(j)} \prod_{k=1}^{j-1}\cos^2\theta_i^{(k)} ~
\end{equation}

Minimizing $E$ requires finding the optimal values of the parameters $m^{(j)}$ and $\theta_i^{(j)}$. However, $E$ depends on infinitely many parameters and optimizing them using numerical methods is impossible. A modified version of generic search with a finite number of parameters can be used to estimate the minimum of expected number of iterations.

\subsection{Modified Method \label{subsec:2.3}}

The procedure is slightly changed to keep the number of parameters finite. In this modified version, maximum number of steps is set to be $n$. If the solution is not found in $n-1$ steps, in the $n^{th}$ step, $m^{(n)}=(\pi/4)\sqrt{N}$ Grover iterations are applied to the initial state of the original Grover algorithm (which ensures finding the solution). 

The expected number of iterations $E$ for this method is very similar to the original one. The main difference is the number of steps. The equation for $E$ for the modified method is
\begin{equation}
	E=\sum_{i=0}^{N-1} p_i \sum_{j=1}^{n}m^{(j)} \prod_{k=1}^{j-1}\cos^2\theta_i^{(k)} ~.
	\label{expected}
\end{equation} 
The parameters of the last step are predetermined so, it leaves $n-1$ steps to optimize.

\subsection{Constraint Equations \label{subsec:2.4}}

The parameters of the method are not independent from each other as it can be seen in Subsection~\ref{subsec:2.1}. A relation between $m^{(j)}$ and $\theta_i^{(j)}$ can be found using Eq.~\eqref{restoreci}. For $N\gg1$, it can be assumed that $c_i^{(j)}$ are very small and that $\arcsin c_i^{(j)} \approx c_i^{(j)}$. By also assuming that $m^{(j)}$ are $O(\sqrt{N})$, Eq.~\eqref{restoreci} can be approximated as
\begin{equation}
    \theta_i^{(j)} \approx 2m^{(j)}c_i^{(j)} ~.
\end{equation}

By taking the square of both sides in this equation, summing with respect to the index $i$ and using the normalization condition, we obtain the constraint equation below
\begin{equation}
	\sum_{i=0}^{N-1} \left[ \theta_i^{(j)} \right] ^2=4\left[m^{(j)}\right]^2 ~.
	\label{const}
\end{equation}
For generic search, there are infinitely many constraint equations and for the modified method, there are $n-1$ of them.

\section{\label{sec:level3}Optimality Conditions}

The optimum values for the modified method can be found, since there are only a finite number of parameters. To do this, equations of optimality are derived using Lagrange multipliers method. To define a function for minimizing the expected number of iterations, the definition of $E$ Eq.~\eqref{expected}, and the constraint equations Eq.~\eqref{const} can be used. The resulting function is

\begin{equation}
	\mathcal{L}=\sum_{i=0}^{N-1} p_i \sum_{j=1}^{n}m_j \prod_{k=1}^{j-1}\cos^2\theta_i^{(k)}-\sum_{j=1}^{n-1}\lambda^{(j)} \Gamma^{(j)} ~,
\end{equation} 
\\
where parameter $\lambda^{(j)}$ is the Lagrange multiplier for the $j^{th}$ constraint and

\begin{equation}
	\Gamma^{(j)}=\sum_{i=0}^{N-1} \left[ \theta_i^{(j)} \right]^2 -4\left[m^{(j)}\right]^2 ~.
\end{equation}

At the optimal point, the variation of $\mathcal{L}$ is equal to zero. For its variation to be equal to zero, the partial derivative of $\mathcal{L}$ with respect to each parameter must be equal to zero. This can be shown as

\begin{equation}
	\delta \mathcal{L}=\frac{d\mathcal{L}}{dm^{(j)}}\delta m^{(j)}+\frac{d\mathcal{L}}{d\theta_i^{(j)}}\delta \theta_i^{(j)}+\frac{d\mathcal{L}}{d\lambda^{(j)}}\delta \lambda^{(j)} ~.
\end{equation}

By equating the partial derivatives to zero, three optimality conditions are found. These are equations of the optimal values of the parameters $m^{(j)}$, $\theta_i^{(j)}$ and $\lambda^{(j)}$; which are denoted as $m^{(j)*}$, $\theta_i^{(j)*}$ and $\lambda^{(j)*}$ respectively. These optimality conditions are written as

\begin{equation}
	\sum_{i=0}^{N-1}p_i\prod_{k=1}^{j-1}\cos^2\theta_i^{(j)*} =8\lambda^{(j)*} m^{(j)*} ~,
	\label{opt1}
\end{equation}
\begin{equation}
	\sum_{i=0}^{N-1}\left[ \theta_i^{(j)*}\right] ^2 =4 \left[m^{(j)*}\right]^2 ~,
	\label{opt2}
\end{equation}
\begin{equation}
	p_i \cos\theta_i^{(\ell)*}\sin\theta_i^{(\ell)*}\sum_{j=\ell+1}^{n}m^{(j)*}\prod\limits_{\substack{k<j \\ k\neq \ell}}\cos^2\theta_i^{(k)*}=\lambda^{(\ell)*} \theta_i^{(\ell)*} ~.
	\label{opt3}
\end{equation}

Solving these equations gives the optimal values for the parameters. Using these values, the minimum of the expected number of iterations can be found. Unfortunately, finding analytical solutions to this set of solutions does not seem possible. Instead, we estimated the optimum values of the parameters using numerical methods. The procedures of these methods and the results of the analyses are given in the next section.

\section{\label{sec:level4}Numerical Analysis}

\subsection{Procedure \label{subsec:4.1}}

In the numerical analyses, the optimality equations for the modified method are used to estimate the parameters. In each analysis, number of steps is set to $n=10$ and the size of the search space is set to $N=10^6$. These numbers are thought to be sufficient to find a close estimation to the original method for big search spaces. In any case, applying a method with infinitely many parameters is not applicable in reality. Thus, finding the minimum for the original method would only determine the lower bound for the expected number of iterations. The values found in the analyses are considered as estimations of these limit values.

Optimality conditions given in Eq.~\eqref{opt1}, Eq.~\eqref{opt2} and Eq.~\eqref{opt3} are put to use for the analyses. Initial values for $\lambda^{(j)}$ are set arbitrarily, and then enhanced iteratively. In this procedure, Newton-Raphson method is employed to find the inverse of some transcendental function which is derived from the optimality equations. After updating $\lambda^{(j)}$ values for all $j$ values, which means at the end of each iteration $m^{(j)}$, $\theta_i^{(j)}$ and $E$ values are calculated. Iteration is halted after the difference between two successive $E$ values is less than a predetermined precision value. The details of this numerical analysis can be seen in \cite{ref13}.

Analyses are made for eight probability distributions depicting power law, exponential and Gaussian behavior. To obtain a discrete probability distribution from a continuous one, it is assumed that the continuous distribution function $f(x)$ is a probability distribution on $0\leq x\leq 1$ interval and we defined
\begin{equation}
	p_i=\int_{i/N}^{(i+1)/N}f(x)dx~, \quad (\textrm{for } i=0,1,\ldots,N-1) ~.
\end{equation}

For power law distributions, $f(x)=(n+1)x^n$, ($n=0,1,\ldots,5)$. For exponential distribution we have $f(x)=Ae^{-cx}$ where $c$ is a large value (we have used $c=30$) and $A$ is for normalization. For half-normal distribution we have $f(x)=Ae^{-cx^2}$ where we have taken $c$ to be $18$. Since the algorithm does not depend on how the elements are ordered, the fact that whether $f(x)$ is increasing or decreasing is irrelevant. However, there is a relation between the standard deviation and $E$ for monotonic distributions. Thus, we gave the results for monotonic distributions in the next section. This is also the reason why only half of the normal distribution is used. If a normal distribution with a mean inside the $(0,1)$ interval were used, identical results would have been obtained with the half-normal ones. We also made the same analyses for the randomized versions of the chosen distributions to verify that ordering of the elements does not change the expected number of iterations. These randomized versions are random permutations of the given ordered distributions.

\subsection{Results \label{subsec:4.2}}

We estimated the optimal values of the parameters for each selected probability distribution. Using these values, we calculated the minimum of the expected number of Grover iterations. These $E$ values are of order $O(\sqrt{N})$. We also calculated the standard deviations $\sigma$ which are $O(N)$. To make the results of our study independent of the size of a search set, we also calculated $E/\sqrt{\sigma}$ values in an effort to find a constant of the method. These results and the improvements against $\pi/4 \sqrt{N}$, for all the distributions are given in Table~\ref{table1}.

\begin{table}[h]
	\caption{The results of the numerical analysis for chosen probability distributions}
	\label{table1}
	\begin{ruledtabular}
        \begin{tabular}{cccc}
            \textrm{$f(x)$}&
            \textrm{$E/\sqrt{N}$}&
            \textrm{$E/\sqrt{\sigma}$}&
            \textrm{Improvement}\\
            \colrule
            $1$ & $0.690$ & $1.285$ & $12\,\%$  \\ 
				
			$2x$ & $0.627$ & $1.292$ & $20\,\%$ \\ 
				
			$3x^2$ & $0.559$ & $1.270$ & $29\,\%$ \\ 
				
			$4x^3$ & $0.507$ & $1.254$ & $35\,\%$ \\ 
				
			$5x^4$ & $0.466$ & $1.241$ & $41\,\%$ \\
				 
			$6x^5$ & $0.433$ & $1.232$ & $45\,\%$ \\ 
				
			exp & $0.213$\footnote{This result is valid for exponential distribution with c=30, but not for every exponential distribution.} & $1.168$ & $73\,\%$\footnote{The amount of improvement is calculated for the exponential distribution with c=30.} \\ 
				
			hnorm & $0.406$\footnote{This result is valid for the half-normal distribution with c=18, but not for every half-normal distribution} & $1.281$ & $48\,\%$\footnote{The amount of improvement is calculated for the half-normal distribution with c=18.} \\
        \end{tabular}
    \end{ruledtabular}
\end{table}

The first entry in the table belongs to the uniform distribution. For this distribution, there is no prior knowledge, thus, no improvement on the results of \cite{ref11} are obtained as expected. However, for the distributions with lower standard deviation values, lower $E$ values are found. For these distributions the decrease in the expected number iterations is more than $12\,\%$. The highest improvement, which is $73\,\%$, is achieved for the exponential distribution.

\begin{figure} [h]
	\centering
	\begin{tikzpicture} [scale=1]
		\draw[-, line width=1.2] (0,0) -- (7.5,0);
		\draw[->, line width=1.2] (0,0) -- (0,4);
		\draw[line width=1.2] (0.8,0) -- (0.8,-0.15);
		\draw[line width=1.2] (1.6,0) -- (1.6,-0.15);
		\draw[line width=1.2] (2.4,0) -- (2.4,-0.15);
		\draw[line width=1.2] (3.2,0) -- (3.2,-0.15);
		\draw[line width=1.2] (4,0) -- (4,-0.15);
		\draw[line width=1.2] (4.8,0) -- (4.8,-0.15);
		\draw[line width=1.2] (5.6,0) -- (5.6,-0.15);
		\draw[line width=1.2] (6.4,0) -- (6.4,-0.15);
		\draw[line width=1.2] (0,2.84) -- (-0.15,2.84);
		\draw[line width=1.2] (0,0.5) -- (-0.15,0.5);
		\draw[line width=1.2] (0,3.5) -- (-0.15,3.5);
		\node at (-0.65,2.84) {$1.267$};
		\node at (-0.65,0.5) {$1.150$};
		\node at (-0.65,3.5) {$1.300$};
		\node at (0.8, 3.194)[circle,fill,inner sep=1.5pt]{};
		\node at (1.6, 3.333)[circle,fill,inner sep=1.5pt]{};
		\node at (2.4, 2.909)[circle,fill,inner sep=1.5pt]{};
		\node at (3.2, 2.573)[circle,fill,inner sep=1.5pt]{};
		\node at (4, 2.323)[circle,fill,inner sep=1.5pt]{};
		\node at (4.8, 2.132)[circle,fill,inner sep=1.5pt]{};
		\node at (5.6, 0.858)[circle,fill,inner sep=1.5pt]{};
		\node at (6.4, 3.123)[circle,fill,inner sep=1.5pt]{};
		\node at (0.8,-0.3){$1$};
		\node at (1.6,-0.3){$2x$};
		\node at (2.4,-0.3){$3x^2$};
		\node at (3.2,-0.3){$4x^3$};
		\node at (4,-0.3){$5x^4$};
		\node at (4.8,-0.3){$6x^5$};
		\node at (5.55,-0.37){$exp$};
		\node at (6.45,-0.31){$hnorm$};
		\node at (0,4.3){$E/\sqrt{\sigma}$};
		\draw[line width=0.5] (0,2.84) -- (7.3,2.84);
		\draw[dashed, line width=0.5] (0,0.5) -- (7.3,0.5);
		\draw[dashed, line width=0.5] (0,3.5) -- (7.3,3.5);
	\end{tikzpicture}	
	\caption{The fitting of the minimums of the expected number of iterations}
	\label{fig1}
\end{figure}
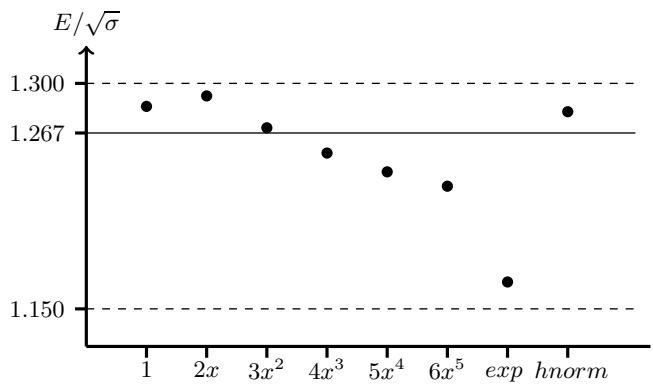

These results suggests that generic search can find the solution in a set with prior knowledge faster than its predecessors. They also suggest that, for monotonic probability distributions, there might be a linear relationship between the square root of the standard deviation and the expected number iterations to be used in the generic search. Using the results given in \ref{table1} and the least square errors method, this linear relationship is estimated as

\begin{equation}
	E\approx1.267\sqrt{\sigma} ~.
	\label{linear}
\end{equation}
How well our results fit in this relation is shown in Figure~\ref{fig1}.

The analyses conducted on the randomized versions of the given probability distributions have verified that the expected number of Grover iterations does not depend on the order of the elements. For each of the distributions, the estimated minimum of $E$ is the same for the ordered and randomized versions. It also shows that Eq.~\eqref{linear} can be used to estimate the minimum of $E$ for any distribution, not only the monotonic ones. However, for non-monotonic distributions, the standard deviation of the distribution itself should not be inserted into the equation. Instead, the standard deviation of the ordered version of the distribution must be used.

\section{\label{sec:level5}Conclusion}

We have shown that generic search method provides an improvement in the expected number of iterations for search spaces with non-uniform probability distributions. We also derived an equation that can be used to find an approximation to the expected number of iterations for any distribution. We also show that generic search is suitable to be used for non-monotonic distributions, aside from the monotonic ones.

In this paper, we assumed that any initial state can be prepared without much complexity. However, some of the initial states to be used to achieve a minimum expected number of iterations might be hard-to-prepare. Computational cost of preparing such quantum states might outweigh that of applying Grover iterations. In that case, generic search does not provide any advantage in the practical applications. For practical use, some other methods that are similar to the one introduced here but use easy-to-prepare quantum states can be designed. Examples of such methods are given in \cite{ref13}. Obviously, these methods does not ensure improvements as big as the one ensured by generic search, but some other methods giving closer results to it can still be developed. Thus, the results that are given for the generic search shows the theoretical minimum of the expected number of iterations for any method employing generic initial states.

\bibliography{main}

\end{document}